\documentclass[a4paper,fleqn,usenatbib]{mnras}

\usepackage{mathptmx}

\usepackage[T1]{fontenc}
\usepackage{ae,aecompl}

\usepackage{graphicx}	
\usepackage{amsmath}	
\usepackage{amssymb}	
\usepackage{bm}

\newcommand{\Lcs}{\bm{L}_{\rm cs}}
\newcommand{\Lc}{\bm{L}_{\rm c}}
\newcommand{\Lcb}{\bm{L}_{\rm cb}}
\newcommand{\Lcsu}{\hat{L}_{\rm cs}}
\newcommand{\Lcu}{\hat{L}_{\rm c}}
\newcommand{\Lcbu}{\hat{L}_{\rm cb}}

\title[Misaligned Binary Discs]{Generating large misalignments in gapped and binary discs}

\author[Owen, J. E. \& Lai, D]{
James E. Owen$^{1}$\thanks{E-mail: jowen@ias.edu}\thanks{Hubble Fellow}
and Dong Lai$^{2,1}$
\\
$^{1}$Institute for Advanced Study, Einstein Drive, Princeton NJ, 08540, USA\\
$^{2}$Center for Astrophysics and Planetary Science, Department of Astronomy, Cornell University, Ithaca, NY 14853, USA
}

\pubyear{2015}

\begin{document}
\label{firstpage}
\pagerange{\pageref{firstpage}--\pageref{lastpage}}
\maketitle

\begin{abstract}
Many protostellar gapped and binary discs show misalignments between their inner and outer discs; in some cases, $\sim70$ degree misalignments have been observed. Here we show that these
misalignments can be generated through a ``secular precession
resonance'' between the nodal precession of the inner disc and the
precession of the gap-opening (stellar or massive planetary) companion. An
evolving protostellar system may naturally cross this resonance during
its lifetime due to disc dissipation and/or companion migration. 
If resonance crossing occurs on the right timescale, of order a few Myrs, characteristic for young protostellar systems, the inner and outer discs can become highly misaligned ($\gtrsim 60$ degrees). When the primary star has a mass of order a solar mass, generating a significant misalignment typically requires the companion to have a mass of $\sim 0.01-0.1$ M$_\odot$ and an orbital separation of tens of AU. The recently observed companion in the cavity of the gapped, highly misaligned system HD~142527 satisfies these requirements, indicating that a previous resonance crossing event misaligned the inner and outer discs. Our scenario for HD~142527's misaligned discs predicts that the companion's orbital plane is aligned with the outer disc's; this prediction should be testable with future observations as the companion's orbit is mapped out. Misalignments observed in several other gapped disc systems could be generated by the same secular resonance mechanism.
\end{abstract}

\begin{keywords}
accretion, accretion disks --- protoplanetary disks ---  binaries: general --- stars: individual (HD 142527) 
\end{keywords}



\section{Introduction}

Newly formed binary stars often undergo a period of co-evolution with
a circumstellar and circumbinary disc system.  In this scenario, a
circumbinary disc is disrupted into accreting streams by the binary,
and feeds onto circumstellar discs around individual stars
\citep[e.g.][]{Shi2012,DOrazio2013,Munoz2016}. Young binary systems
with both circumstellar and circumbinary 
discs 
have been observed for some time \citep[e.g.][]{Andersen1989,Mathieu1997}.
Interestingly, recent imaging observations with high angular resolution have revealed 
that the circumstellar and circumbinary discs are often misaligned,
rather than sharing a common orbital plane.

In a picture where binaries form from an isolated
rotating molecular cloud core, strongly misaligned discs are rather
strange where one would expect everything to be aligned. Furthermore,
if the binary companion forms through disc fragmentation
\citep[e.g.][]{Adams1989,Rice2005,Matzner2005,Kratter2006,Clarke2009,Meru2011},
then one would naturally expect it to share the same orbital plane as
the disc from which it formed. Simulations of turbulent star
formation, where the cores do not appear to be isolated from the
surrounding molecular clouds, indicate that accretion from different
directions can cause the angular momentum vector of the disc to evolve
in single star systems \citep[e.g.][]{Bate2010}, and a similar process
is expected to happen in binary systems.
Accretion through the binary/disc is generally expected to drive the
system towards alignment \citep{Foucart2013}, although small
misalignments are possible \citep{Foucart2014}.

There are several observations of young binary systems that indicate
misalignments between the circumbinary disc, circumstellar disc and
the binary orbital plane
\citep[e.g.][]{Winn2004,Hioki2011,Kennedy2012,Avenhaus2014,Brinch2016,Kraus2017,Takakuwa2017}.
Additionally, several protoplanetary discs with large gaps/cavities 
(often called ``transition discs'' -- see \citealt{Casassus2016,Owen2016} for recent reviews), 
which could contain massive companions, are observed to contain misaligned inner and outer discs.
Of particular interest is the system HD~142527, which contains an
extended, massive disc (hundreds of AUs, $\sim 0.1M_\odot$) with a
large cleared dust gap from $\sim 10$~AU to $\sim 100$~AU \citep{Fukagawa2006}; an
accreting M-dwarf companion (with mass about 10-20 times smaller than
the 2~M$_\odot$ Herbig Ae/Be primary) has been found to orbit within
the cavity \citep{Biller2012,Lacour2016}.  Recent NIR scattered light observations
\citep[e.g.][]{Avenhaus2014} have revealed notches and shadows,
indicating that the inner disc and the outer disc are misaligned by
$\sim$70 degrees \citep{Marino2015} -- This misalignment is confirmed
by the gas kinematics from resolved CO observations
\citep{Casassus2015a}.  Another system, HD~100453, shows similar
features to HD~142527 in scattered light observations, and have been
interpreted as a misalignment between the inner and outer discs of 72
degrees \citep{Benisty2016}.
Yet another system, HD~135344B, shows features in its image
that indicate a weaker misalignment between its
inner and outer disc of 22 degrees \citep{Stolker2016}.

In this paper, we explore a dynamical mechanism to generate large
misalignments in gapped protoplanetary discs starting from a nearly
co-planar configuration.  The proto-typical system we study consists of an
inner disc and an outer disc, both assumed to be nearly flat (see
Section 2), with a gap produced by a low-mass binary companion. Such a setup includes ``transition'' discs if they do contain massive companions. The gravitational interaction between the discs and the companion
generates mutual nodal precession of the disc's and binary's orbital planes.
As the system evolves in time (e.g., due to accretion or dissipation of the discs),
these precession frequencies match and may lead to inclination excitation.
Such ``secular precession resonance'' has long been studied in the context of 
planetary spin dynamics \citep[e.g.][]{Ward1973}.
Recently, Batygin \& Adams (2013) considered the effect of secular
precession resonance as a mechanism of generating misalignment between the
stellar spin and protoplanetary disc in the presence of an external
binary companion.  Lai (2014) presented a simple way (based on vector
equations) of studying this resonance during  the disc evolution and also
included the effects accretion and magnetic fields (see also Spalding
\& Batygin 2014). Matsakos \& Konigl (2017) studied a variant of this
``primordial disc misalignment'' model in which the binary companion
is a giant planet orbiting in the gap between the inner and outer disc.

Our motivation in this paper is the misaligned disc systems like HD~142527
and HD~100453.  Our system is similar to that studied by Matsakos \&
Konigl (2017); however, we ignore the stellar spin (since it has a
negligible angular momentum compared to the discs and the embedded
companion), and we focus on generating misalignments between the
circumsteller (inner) and circumbinary (outer) discs, rather than on
misalignments of the star's spin.
The basic setup is described in Section~2, and the concept of secular
precession resonance is discussed further in Section~3. In Section~4 we
numerically investigate various possible scenarios. In Section~5 we put
forward a simple qualitative understanding of our results and
discuss them in the context of observations. Finally, we summarise our
findings in Section~6.

\section{Problem Description}

We consider a primary star of mass $M_*$ surrounded by a circumstellar disc of mass $M_{\rm cs}$ and radial extent $R_{\rm cs}$. The primary is then orbited by a lower-mass companion with mass $M_{\rm c}$ and separation $a_{\rm c}$. The companion could be a lower-mass star, brown dwarf or giant planet. The primary star and companion are then surrounded by a circumbinary disc of mass $M_{\rm cb}$ and radial extent from $R_{\rm in}$ to $R_{\rm out}$. Similar to previous works \citep{Batygin2013,Lai2014,Matsakos2016}, we take the two discs to be flat, each having a single orientation. This assumption is reasonable because different regions of the disc can communicate with each other efficiently by internal stresses (such as bending waves\footnote{We note protostellar discs are in the ``wave-like'' regime where bending waves propagate with speeds of order half the sound speed, rather than the diffusive regime \citep[e.g.][]{Papaloizou1995}}), so that any disc warp is small \citep{Foucart2014}. Thus, our system is specified by three vectors: the angular momentum of the circumstellar disc ($\Lcs$), the angular momentum of the companion ($\Lc$) and the angular momentum of the circumbinary disc ($\Lcb$). There are six torques between the various components of which three are independent. We calculate the torques using the quadrupole approximation and assume that the primary star is much more massive than the companion and the discs \citep[e.g.][]{Lai2014,Tremaine2014}.

The torque on the circumstellar disc due to the companion (${\bm \tau}_{\rm cs,c}$) is:
\begin{equation}
{\bm \tau}_{\rm cs,c}=\frac{3GM_{\rm c}}{4a_{\rm c}^3}\left(\Lcsu\cdot\Lcu\right)\Lcsu\times\Lcu\int_{0}^{R_{cs}}R^2dM_{cs}\label{eqn:taucsc}
\end{equation}
where $\hat{L_i}$ are unit vectors. To get a sense of the magnitude and form of this torque we can crudely evaluate the integral (assuming, as in most protostellar discs that the disc mass is dominated at large radius, but without specifically specifying its form -- we will do this later in Section~\ref{sec:details}) to find (neglecting the $(\Lcsu\cdot\Lcu)\Lcsu\times\Lcu$ factor):
\begin{equation}
|{\bm \tau}_{\rm cs,c}|\sim \frac{GM_{\rm c}M_{\rm cs}R_{\rm cs}^2}{a_{\rm c}^3}
\end{equation}
The torque on the companion due to the circumbinary disc ($\tau_{\rm c,cb}$) is:
\begin{equation}
{\bm \tau}_{\rm c,cb}=\frac{3GM_{\rm c}a_{\rm c}^2}{4}\left(\Lcu\cdot\Lcbu\right)\Lcu\times\Lcbu\int_{R_{\rm in}}^{R_{\rm out}}\frac{dM_{\rm cb}}{R^3}\label{eqn:tauccb}
\end{equation}
Similiary the magnitdue of this torque is approximately:
\begin{equation}
|{\bm \tau}_{\rm c,cb}|\sim \frac{GM_{\rm c}M_{\rm cb}^La_{\rm c}^2}{R_{\rm in}^3}
\end{equation}
where $M_{\rm cb}^L$ is the total disc mass ``locally'' around $R_{\rm in}$, specifically $M_{\rm cb}^L\equiv2\pi R_{\rm in}^2\Sigma(\rm R_{\rm in})$.
Finally the torque on the circumstellar disc due to the circumbinary disc $(\tau_{\rm cs,cb})$ is:
\begin{equation}
{\bm \tau}_{\rm cs,cb}= \int_0^{R_{\rm cs}}R^2dM_{\rm cs}\int_{R_{\rm in}}^{R_{\rm out}}\frac{3GdM_{\rm cb}}{4R'^3}\left(\Lcsu\cdot\Lcbu\right)\Lcsu\times\Lcbu\label{eqn:taucscb}
\end{equation}
which has a magnitude of order:
\begin{equation}
|{\bm \tau}_{\rm cs,cb}|\sim\frac{GM_{\rm cs}M_{\rm cb}^LR_{cs}^2}{R_{\rm in}^3}
\end{equation}
With these torques the evolution of the angular momenta in the system can be written:
\begin{eqnarray}
\frac{{\rm d}\Lcs}{{\rm d}t}&=&{\bm \tau}_{\rm cs,c}+{\bm \tau}_{\rm cs,cb}+\frac{{\rm d}|\Lcs|}{{\rm d}t}\Lcsu\label{eqn:Lcs_e}\\
\frac{{\rm d}\Lc}{{\rm d}t}&=&-{\bm \tau}_{\rm cs,c}+{\bm \tau}_{\rm c,cb}+\frac{{\rm d}|\Lc|}{{\rm d}t}\Lcu\label{eqn:Lc_e}\\
\frac{{\rm d}\Lcb}{{\rm d}t}&=&-{\bm \tau}_{\rm cs,cb}-{\bm \tau}_{\rm c,cb}+\frac{{\rm d}|\Lcb|}{{\rm d}t}\Lcbu\label{eqn:Lcb_e}
\end{eqnarray}
This system contains many different precession frequencies allowing for the possibility of a ``secular precession resonance'' that could result in large misalignments between the two discs. We note, as we will discuss later, a resonance does not necessarily result in a large misalignment; in these models a resonance is a necessary, but not a sufficient condition. 

\subsection{Precession Frequencies}\label{sec:details}

To evaluate the precession frequencies, we consider a power-law surface density profile of the form:
\begin{equation}
\Sigma=\Sigma_0\left(\frac{R}{R_0}\right)^{-1}
\end{equation}
Such a choice is  the typical surface density profile obtained from mm-images of protoplanetary discs \citep[e.g.][]{Andrews2009,Andrews2010} and is expected for a constant `alpha-disc' model, with a flared structure (e.g. \citealt{Hartmann1998}, where temperature $\propto R^{-1/2}$ -- \citealt{Kenyon1987}).  With our choice of density profile, the angular momenta of the circumstellar and circumbinary discs are $2/3M_{\rm cs}\Omega_K(R_{\rm cs})R_{\rm cs}^2$ and $2/3M_{\rm cb}\Omega_K(R_{\rm cb})R_{\rm cb}^2$ respectively (with $\Omega_K(R)$ the Keplerian angular velocity at radius $R$). The precession frequencies are defined via ${\bm \tau}_{\rm i,j}=\Omega_{i,j}\hat{L}_j\times{\bm L}_i$, and note that $\Omega_{j,i}=\Omega_{i,j}|{\bm L}_i|/|{\bm L}_j|$. The six individual precession frequencies are:

\begin{eqnarray}
\Omega_{\rm cs,c}&=&-\frac{3}{8}\,\Omega_K\left(R_{\rm cs}\right)\left(\frac{M_{\rm c}}{M_*}\right)\left(\frac{R_{\rm cs}}{a_{\rm c}}\right)^3\Lcsu\cdot\Lcu\label{eqn:Ocsc}\\
\Omega_{\rm cs,cb}&=&-\frac{3}{16}\,\Omega_K\left(R_{\rm cs}\right)\left(\frac{M_{\rm cb}}{M_*}\right)\left(\frac{R_{\rm cs}^3}{R_{\rm out}R_{\rm in}^2}\right)\Lcsu\cdot\Lcbu\\
\Omega_{\rm c,cb}&=&-\frac{3}{8}\,\Omega_K\left(a_{\rm c}\right)\left(\frac{M_{\rm cb}}{M_*}\right)\left(\frac{a_{\rm c}^3}{R_{\rm out}R_{\rm in}^2}\right)\Lcu\cdot\Lcbu\\
\Omega_{\rm c,cs}&=&-\frac{1}{4}\,\Omega_K\left(a_{\rm c}\right)\left(\frac{M_{\rm cs}}{M_*}\right)\left(\frac{R_{\rm cs}}{a_{\rm c}}\right)^2\Lcu\cdot\Lcsu\\
\Omega_{\rm cb,c}&=&-\frac{3}{4}\,\Omega_K\left(R_{\rm out}\right)\left(\frac{M_{\rm c}}{M_*}\right)\left(\frac{a_{\rm c}}{R_{\rm in}}\right)^2\Lcbu\cdot\Lcu\\
\Omega_{\rm cb,cs}&=&-\frac{3}{16}\,\Omega_K\left(R_{\rm out}\right)\left(\frac{M_{\rm cs}}{M_*}\right)\left(\frac{R_{\rm cs}}{R_{\rm in}}\right)^2\Lcbu\cdot\Lcsu
\end{eqnarray} 
For systems that are close to being aligned initially there are  three possible precession resonances: a resonance between the circumstellar disc and the companion ($\Omega_{\rm cs}=\Omega_{\rm c}$, where $\Omega_{\rm cs}=\Omega_{\rm cs,c}+\Omega_{\rm cs,cb}$ and $\Omega_{\rm c}=\Omega_{\rm c,cs}+\Omega_{\rm c,cb}$); a resonance between the circumstellar disc and the circumbinary disc ($\Omega_{\rm cs}=\Omega_{\rm cb}$, where $\Omega_{\rm cb}=\Omega_{\rm cb,cs}+\Omega_{\rm cb,c}$) and finally a resonance between the companion and the circumbinary disc ($\Omega_{\rm c}=\Omega_{\rm cb}$). Not all these resonances can occur in realistic systems or lead to large misalignments as we shall demonstrate below.  

\section{Secular Precession Resonances}

 To assess the importance of various resonances we consider a set of specific examples before performing numerical integrations. In each example we assume one component of the system has so much angular momentum that it precesses so slowly that the corresponding angular momentum unit vector can be considered constant.

\subsection{A massive circumbinary disc  ($L_{\rm cb}\gg L_{\rm cs},\,L_{\rm c}$)}\label{sec:massive_outer}
Perhaps the most natural setup consists of a massive extended circumbinary disc that dominates the angular momentum budget of the system. Therefore, only the circumstellar disc and the companion precess, and a secular precession resonance occurs when $\Omega_{\rm cs}=\Omega_{\rm c}$, or: 
\begin{equation}
\Omega_{\rm cs,c}+\Omega_{\rm cs,cb}=\Omega_{\rm c,cb}+\Omega_{\rm c,cs}
\end{equation}
For a massive, large circumbinary disc, $\Omega_{\rm c,cb}/\Omega_{\rm c,cs}\gg 1$, unless the circumbinary disc is abnormally far away from the companion. In the scenario we are envisaging, it is the companion itself that truncates the circumstellar and circumbinary disc, so we would expect the inequality to be readily satisfied. Furthermore, we would also expect $\Omega_{\rm cs,c}/\Omega_{\rm cs,cb}\gg1$, as to truncate the circumstellar disc, the tidal torque from the binary should dominate over that from the circumbinary disc. Hence, the resonance condition is simplified to:
\begin{equation}
\Omega_{\rm cs,c}\simeq\Omega_{\rm c,cb}\label{eqn:simple_cond1}
\end{equation}
or,
\begin{equation}
\left(\frac{M_{\rm c}}{M_{\rm cb}^L}\right)\sim \left(\frac{a_{\rm c}}{R_{\rm cs}}\right)^{3/2}\left(\frac{a_{\rm c}}{R_{\rm in}}\right)^3\label{eqn:crit_res1}
\end{equation}
When a circumbinary disc accretes on to a companion, starting with $M_{\rm cb}\gg M_c$, to  $M_{\rm cb}\ll M_c$, one would expect the criterion in Equation~\ref{eqn:crit_res1} could be satisfied at some point during the evolution, leading to resonant behaviour and possibly large changes in the alignment angles. Setting Equation~\ref{eqn:crit_res1} to have parameters similar to those observed in HD~142527 we find:
\begin{eqnarray}
M_{\rm cb}^L&\sim& 2\,{\rm M}_{J} \left(\frac{M_{\rm c}}{0.1\,{\rm M}_\odot}\right)\left(\frac{R_{\rm cs}}{10\,{\rm AU}}\right)^{3/2}\nonumber \\&&\times\left(\frac{a_{\rm c}}{20\,{\rm AU}}\right)^{-9/2}\left(\frac{R_{\rm in}}{100\,{\rm AU}}\right)^3
\end{eqnarray}
Local disc masses of order jupiter masses are observed in many gapped systems \citep[e.g.][]{Andrews2010}. In the case of HD~142527, \citet{Casassus2015b} measure a local disc mass of approximately a Jupiter mass, assuming a gas-to-dust mass ratio of 100. Therefore, we anticipate it is likely a resonance between the precession of the companion and circumstellar disc can occur in real systems.    

To determine whether the resonance (Equation ~\ref{eqn:simple_cond1}) can lead to a significant production of misalignment, we can use a ``geometric picture'' following Lai (2014). In the limit of 
$|\Lcb|\gg |\Lc|\gg|\Lcb|$., the unit angular momentum vectors evolve according to 
\begin{eqnarray}
\frac{{\rm d}\Lcsu}{{\rm d}t}&\approx&\Omega_{\rm cs,c}\,\Lcu\times\Lcsu\label{eqn:init1}\\
\frac{{\rm d}\Lcu} {{\rm d}t}&\approx&\Omega_{\rm c,cb}\,\Lcbu\times\Lcu\label{eqn:init2}
\end{eqnarray}
We can transform into a frame rotating with the precession of the companion at a rate $\Omega_{\rm c,cb}$, such that $\Lcu$ remains fixed in time. In this rotating frame the evolution of $\Lcsu$ is governed by:
\begin{equation}
\left(\frac{{\rm d}\Lcsu}{{\rm d}t}\right)_{\rm rot}\approx\Omega_{\rm cs,c}\,\Lcu\times\Lcsu-\Omega_{\rm c,cb}\Lcbu\times\Lcsu\label{eqn:rot1}
\end{equation}
This tells us that $\Lcsu$ precesses around a vector $\hat{L}_r$ with precession rate $\Omega_r$, given by:
\begin{equation}
\Omega_r\hat{L}_r=\Omega_{\rm cs,c}\Lcu-\Omega_{\rm c,cb}\Lcbu\label{eqn:Or2}
\end{equation}
As the system evolves from being far from resonance with $\Omega_{\rm cs,c}\gg\Omega_{\rm c,cb}$ to  $\Omega_{\rm cs,c}\ll\Omega_{\rm c,cb}$, the circumstellar disc goes from precessing about $\Lcu$ to precessing around $\Lcbu$. However, close to resonance when $\Omega_{\rm cs,c}\sim\Omega_{\rm c,cb}$ the vector $\hat{L}_r$ about which $\Lcsu$ is precessing deviates from $\Lcbu$ by a large angle. This is obvious if one considers the case where $\Lcu$ and $\Lcbu$ are misaligned by some small angle $\theta$, the one finds:
\begin{equation}
\arccos\left(\hat{L}_r\cdot\Lcbu\right)\approx\frac{\pi}{2}+\frac{\theta}{2}
\end{equation}
Thus, in resonance $\Lcsu$ is precessing around an axis that is almost orthogonal to $\Lcbu$.  For a large misalignment to be generated, a necessary (but not sufficient) condition is $\Omega_r^{-1}$ is comparable to, or shorter than the timescale on which the system is evolving. If resonance crossing is too fast, $\Lcsu$ is unable to precess far enough around $\hat{L}_r$ to generate a large misalignment. Finally, it is clear from Equation~\ref{eqn:init2} that the misalignment between the companion and circumbinary disc is constant.

\subsection{A massive companion, ($L_{\rm c}\gg L_{\rm cs},\,L_{\rm cb}$)}\label{sec:massive_companion} 
One can imagine late in the evolution of a protostellar system, when the discs contain very little mass, the companion dominates the total angular momentum budget. Therefore, a ``resonance'' can occur between the precessing circumstellar and circumbinary discs when $\Omega_{\rm cs}=\Omega_{\rm cb}$, or:
\begin{equation}
\Omega_{\rm cs,c}+\Omega_{\rm cs,cb}=\Omega_{\rm cb,c}+\Omega_{\rm cb,cs}\label{eqn:cond_gen2}
\end{equation}
In this case, the massive companion predominately drives the precession of both the inner and outer discs and Equation~\ref{eqn:cond_gen2} reduces to:
\begin{equation}
\Omega_{\rm cs,c}\simeq\Omega_{\rm cb,c}
\end{equation}
or,
\begin{equation}
a_c\sim 70\,{\rm AU}\left(\frac{R_{\rm cs}}{10\,{\rm AU}}\right)^{3/10}\left(\frac{R_{\rm in}}{100\,{\rm AU}}\right)^{2/5}\left(\frac{R_{\rm out}}{300\,{\rm AU}}\right)^{3/10}\label{eqn:crit_res2}
\end{equation}
This criterion is independent of mass and could be satisfied in real systems. In this scenario, the disc angular momentum unit vectors evolve according to:
\begin{eqnarray}
\frac{{\rm d}\Lcsu}{{\rm d}t}&\approx&\Omega_{\rm cs,c}\,\Lcu\times\Lcsu\label{eqn:init4}\\
\frac{{\rm d}\Lcbu}{{\rm d}t}&\approx&\Omega_{\rm cb,cs}\Lcu\times\Lcbu\label{eqn:init6}
\end{eqnarray}
Obviously, both discs are precessing around the same axis ($L_{\rm c}$) {\it independently}.  As such, in ``resonance'' a large misalignment cannot be generated from an initially small misalignment.

\subsection{A massive circumstellar disc, ($L_{\rm cs}\gg L_{\rm c},\,L_{\rm cb}$)}\label{sec:res_crit3}

While a circumstellar disc which contains most of the angular momentum seems strange, it is not so unusual from the perspective of formation of the companion by fragmentation. Fragmentation is expected to occur in the outer regions of massive circumstellar discs \citep[e.g.][]{Clarke2009}. At the point of formation, the angular momentum could still be dominated by the circumstellar disc. The companion is then expected to migrate through the disc to smaller separations \citep[e.g.][]{Li2015,Stamatellos2015}, wherein the more usual hierarchy of a circumbinary disc which contains most of the angular momentum is restored (Section 3.1). Here we assume that the circumstellar disc precesses so slowly that we can consider it to be fixed, thus a resonance can occur when the precession frequencies of the companion and circumbinary disc become equal, $\Omega_{\rm c}=\Omega_{\rm cb}$, or:
\begin{eqnarray}
\Omega_{\rm c,cs}+\Omega_{\rm c,cb}=\Omega_{\rm cb,c}+\Omega_{\rm cb,cs} \label{eqn:cond_gen3}
\end{eqnarray}
In this case of a massive inner disc, $\Omega_{\rm c,cs}/\Omega_{\rm c,cb}\gg 1$. For the precession of the circumbinary disc, it is not clear that either the circumstellar or companion will dominate its precession. Geometric arguments, indicate a strong resonance will occur if $\Omega_{\rm cb,c}/\Omega_{\rm cb,cs}\gg1$. In this case the Equation~\ref{eqn:cond_gen3} simplifies to:
\begin{equation}
\Omega_{\rm c,cs}\simeq\Omega_{\rm cb,c}
\end{equation}
or,
\begin{equation}
M_{\rm cs}\sim 3M_{\rm c}\sqrt{\frac{a_c^3}{R_{\rm out}^3}}\left(\frac{a_c^4}{R_{\rm cs}^2R_{\rm in}^2}\right) \label{eqn:res_crit3}
\end{equation}
Now requiring that the companion forms by fragmentation out of the disc that will make up the circumbinary disc means that $M_{\rm c}\lesssim M_{\rm cb}$. For fragmentation to take place one normally requires the outer regions to have a mass $\sim 0.1$\,M$_*$. Combing all these criteria, along with $M_{\rm cs}\gg M_{\rm cb}$ and $M_{\rm cs}\gg M_{\rm c}$, implies a circumstellar disk more massive than the central star. We conclude that the precession resonance between  the companion and circumbinary disc is unlikely to occur in realistic binary-disc systems. This inference is confirmed by our numerical investigation where we can only get resonant behaviour for  unphysically large circumstellar disc masses.

\section{An evolving system}

We consider two basic scenarios. In the first (Section 4.1) case the companion remains at a fixed separation and we allow the discs to evolve and accrete mass onto the companion. In the second scenario (Section 4.2), we allow the companion to migrate through the disc while the total mass of the two discs remains fixed.  
It is easiest to work in an inertial Cartesian frame, where we initialise the angular momentum unit vectors of the components to:
\begin{eqnarray}
\Lcsu&=&[0,0,1]\\
\Lcu&=&\left[0,-\sin\left(\theta_0\right),\cos\left(\theta_0\right)\right]\\
\Lcbu&=&\left[-\sin\left(\theta_0\right),0,\cos\left(\theta_0\right)\right]
\end{eqnarray} 
where $\theta_0$ is a small angle which we set to 5 degrees. We have checked that our results are qualitatively independent of the initial choice of orientation of the three vectors and the choice of $\theta_0$, provided it is small but non-zero.

\subsection{Evolving Discs}\label{sec:evolve_discs} 
In the ``evolving disc'' scenario we assume the companion remains on a orbit with a fixed separation, but the discs deplete due to  accretion. Simple, self-similar viscous accretion theory \citep{LyndenBell1974,Hartmann1998} suggests that for $\Sigma\propto R^{-1}$ the disc mass ($M_{\rm d}$) declines as:
\begin{equation}
M_{\rm d}=\frac{M_d^0}{\left(1+t/t_\nu\right)^{1/2}} \label{eqn:disc_mass_time}
\end{equation}
where $t_\nu$ is the ``viscous timescale'' that set the time-scale for global disc evolution. Note, this form differs from the temporal evolution of the disc mass used by previous authors \citep[e.g.][]{Batygin2013,Spalding2014,Lai2014,Matsakos2016}. We take Equation~\ref{eqn:disc_mass_time} to specify how the circumbinary disc evolves. In the case where the companion accretes, we assume it accretes some fraction ($f$) of the material that is depleted from the circumbinary disc. Therefore, the circumbinary disc evolves as:
\begin{equation}
M_{\rm cb}=\frac{M_{\rm cb}(t=0)}{\left(1+t/t_\nu\right)^{1/2}}
\end{equation}
and the companion evolves as:
\begin{equation}
M_{\rm c}=M_{\rm c}(t=0)+fM_{\rm cb}(t=0)\left[1-\frac{1}{\left(1+t/t_\nu\right)^{1/2}}\right]
\end{equation}
 The numerical value of $f$ is uncertain; for example it is not clear whether the circumstellar disc will receive most of the material ($f<0.5$) or whether the companion will accrete most of the material ($f>0.5$). Thus, we will keep $f$ as a free parameter.
The remaining mass that does not accrete onto the companion resupplies the inner disc. Since the viscous time of the inner disc is considerably shorter than that of the circumbinary disc, the circumstellar disc can be considered to be in a steady state with a constant mass supply. Where the accretion rate through the circumstellar disc is $\dot{M}_{\rm cs}=(1-f)\dot{M}_{\rm cb}$. As a steady disc has surface density proportional to $\dot{M}_{\rm cs}$, then we can write: 
\begin{equation}
M_{\rm cs}=\frac{M_{\rm cs}(t=0)}{\left(1+t/t_\nu\right)^{3/2}}
\end{equation}   
As the companion grows in mass, the truncation radii of the circumstellar and circumbinary discs due to the companion's torque evolve. For high-mass ratio binaries (with $M_{\rm c}/M_{*}$ not too small), the disc is typically truncated at several Hill radii away from the companion \citep[e.g.][]{Artymowicz1994,Crida2006}. For concreteness, we adopt the following 
prescription for the disc truncation radii:
\begin{eqnarray}
R_{\rm cs}&=&a_c\left[1-3\left(\frac{M_{\rm c}}{3M_*}\right)^{1/3}\right]\label{eqn:trun1}\\
R_{\rm in}&=&a_c\left[1+3\left(\frac{M_{\rm c}}{3M_*}\right)^{1/3}\right]\label{eqn:trun2}
\end{eqnarray}

\subsubsection{Results of numerical integrations}
 
As our canonical example we adopt the parameters which are typical of gapped discs: $M_*=2\,$M$_\odot$, $a_{\rm c}=40$~AU, $M_{\rm c}(t=0)=10^{-3}\,$M$_\odot$, $M_{\rm cs}(t=0)=3\times10^{-3}\,$M$_\odot$, $M_{\rm cb}(t=0)=0.25\,$M$_{\odot}$ and $R_{\rm out}=300\,$AU, with evolution parameters $f=0.5$ and $t_\nu=2\times10^{5}\,$years. In this example the companion will reach a mass of $\sim 0.1\,$M$_\odot$ after 5~Myr of accretion.

\begin{figure}
\centering
\includegraphics[width=\columnwidth]{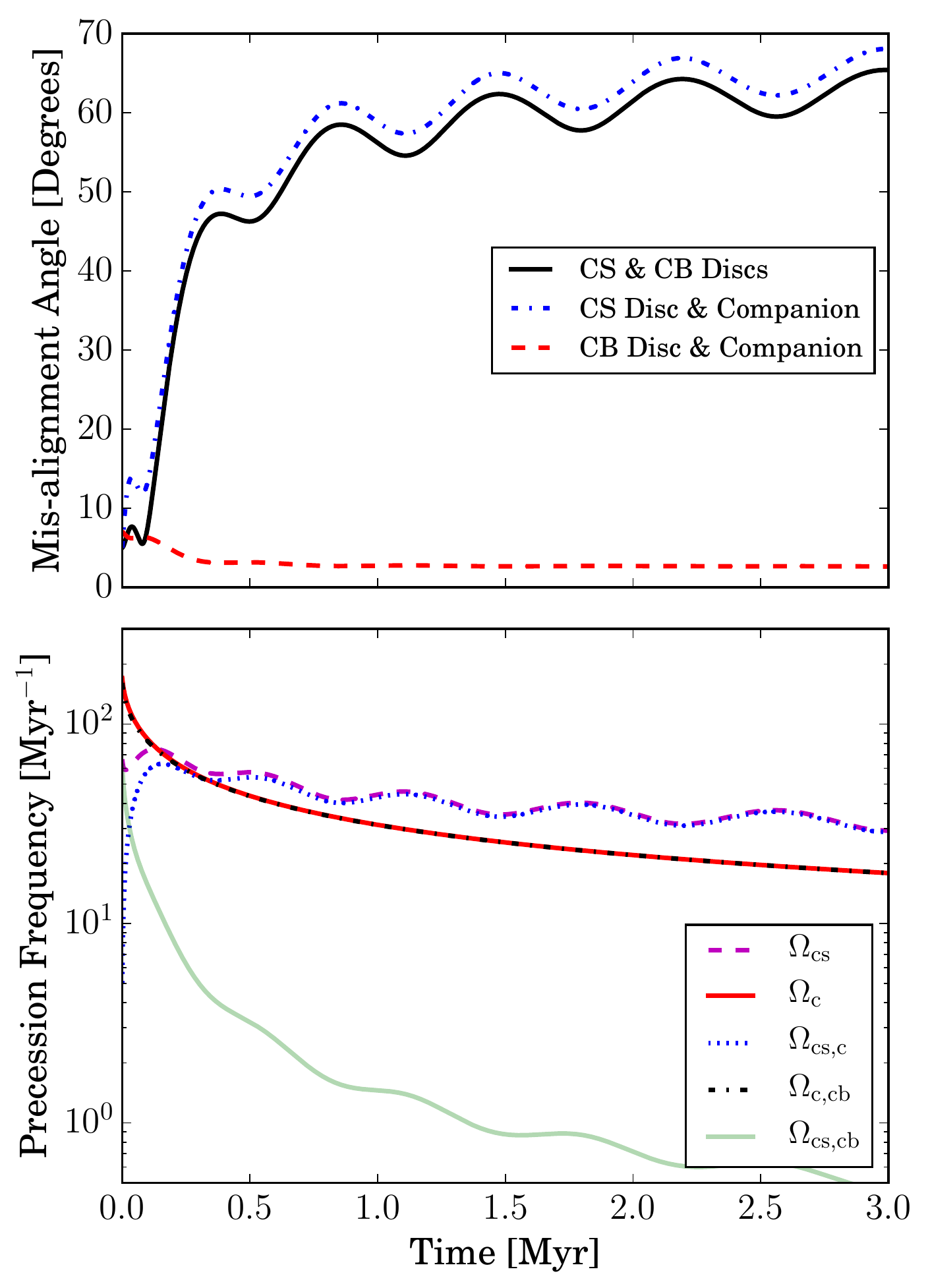}
\caption{The top panel shows the various angles between the angular momentum vectors of the circumstellar disc, circumbinary disc and the companion as a function of time. The bottom panel shows the precession frequencies of the circumstellar disc and the companion as well as the indvidual dominant components. These plots are for the case of an accreting companion and viscously depleting discs with $t_\nu=2\times10^5$~years (see Section 4.1.1).}\label{fig:ac_r1_angle}
\end{figure}  

\begin{figure}
\centering
\includegraphics[width=\columnwidth]{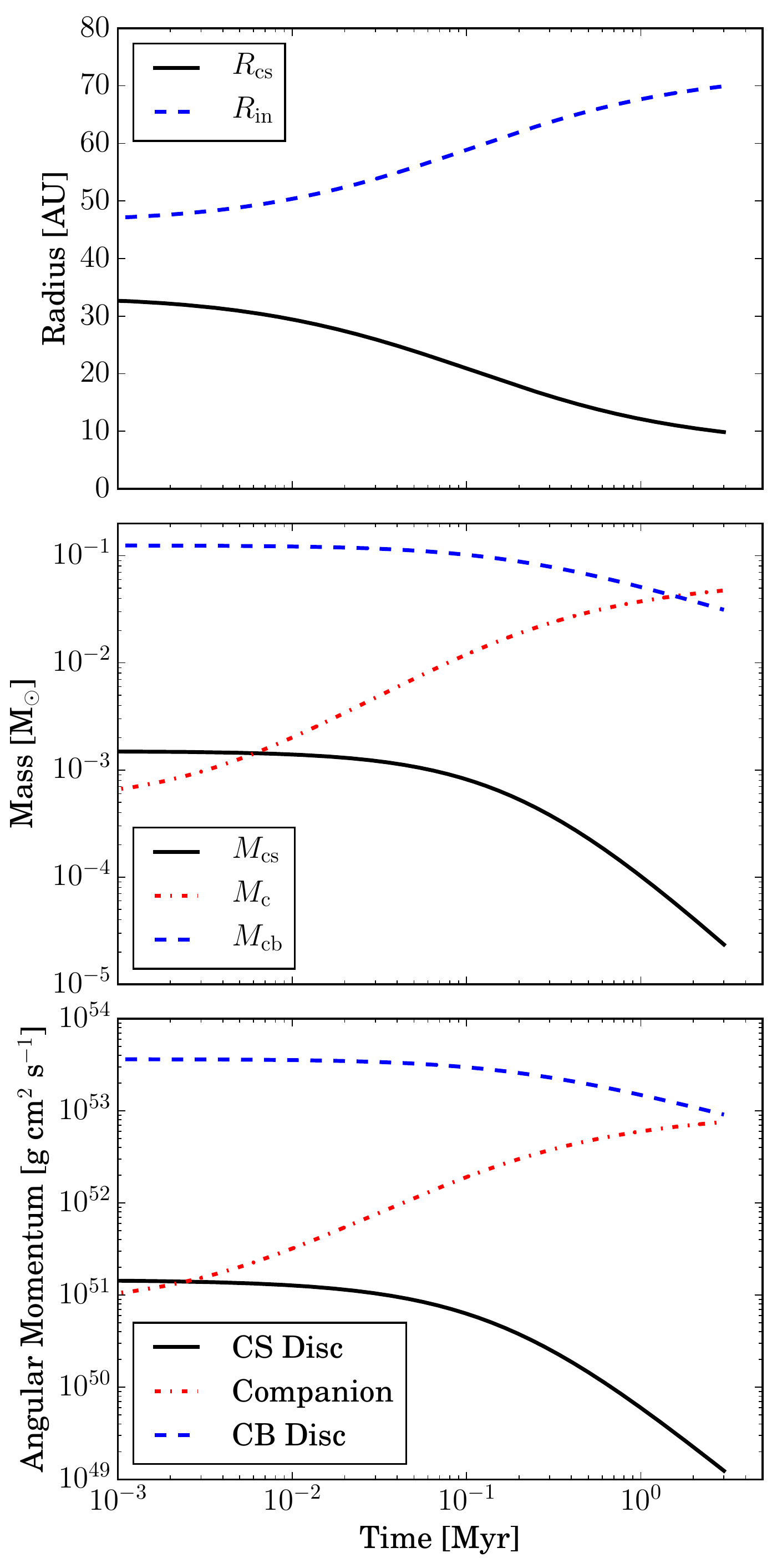}
\caption{Evolution of the disc/companion system as a function of time. The top panel shows the truncation radii of the circumstellar and circubinary discs. The middle panel shows the evolution of the masses of the circumstellar disc, companion and circumbinary disc and finally the bottom panel shows the evolution of the magnitdue of the total angular momenta of the circumstellar disc, companion and circumbinary disc. This evolving system is for the accreting companion scenario shown in Figure~\ref{fig:ac_r1_angle}. }\label{fig:r1_prop}
\end{figure}

In Figure~\ref{fig:ac_r1_angle}, we show the evolution of the mutual inclinations between the discs and companion as well as their precession frequencies. In this scenario, the circumbinary disc contains enough angular momentum such that its precession timescale is long compared to the evolution of the system. In Figure~\ref{fig:r1_prop}, we show the evolution of various systems parameters. We see in Figure~\ref{fig:ac_r1_angle} that the precession frequencies of the circumstellar disc and companion become comparable after about $\sim 0.25$~Myr of evolution. This causes a resonant response where the circumstellar and circumbinary disc strongly misalign, with a misalignment of $\sim 60-70$ degrees found. After the discs have misaligned, the circumstellar disc precesses on a time-scale $\sim 0.5$~Myr with a nutation amplitude of about 10 degrees, driven by the torque from the circumbinary disc. 

As can be seen in Figure~\ref{fig:r1_prop}, the resonance occurs in the case where the outer disc dominates the angular momentum and mass budget of the system, and is representative of the case discussed in Section~\ref{sec:massive_outer}. 

\begin{figure*}
\centering
\includegraphics[width=\textwidth]{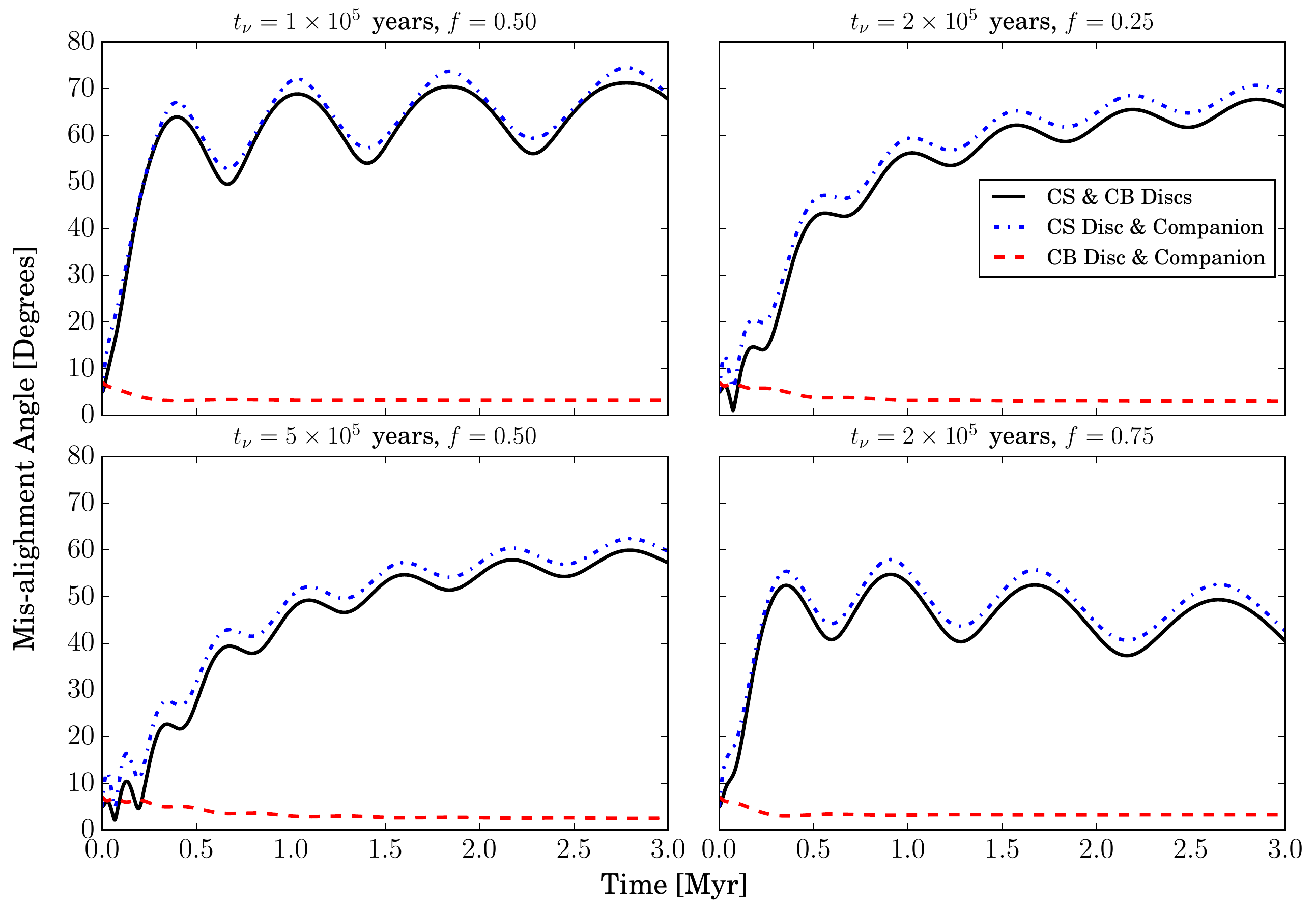}
\caption{The evolution of the misalignment angle for the system shown in Figure~\ref{fig:ac_r1_angle}, but with the viscous time ($t_{\nu}$) varied from $1\times10^{5}$~years to $5\times10^{5}$~years and the accrection efficiency of the compainon ($f$) varied from 0.25 to 0.75.}\label{fig:run1_vary}
\end{figure*}

We can investigate how the evolution parameters effect the results in Figure~\ref{fig:run1_vary}, where we vary the viscous time and $f$.
The evolution curves show that the same evolution is achieved if the disc evolution parameters are varied: rapid evolution towards resonance crossing (either by a shorter viscous time or larger $f$) causes a large rapid change in the misalignment angle, whereas for a slower approach towards resonance crossing results in a smoother change over several precession periods of the circumstellar disc. In all cases we find that resonance crossing can produce large misalignments between the circumstellar and circumbinary disc, while aligning the companion with the circumbinary disc. 

\begin{figure}
\centering
\includegraphics[width=\columnwidth]{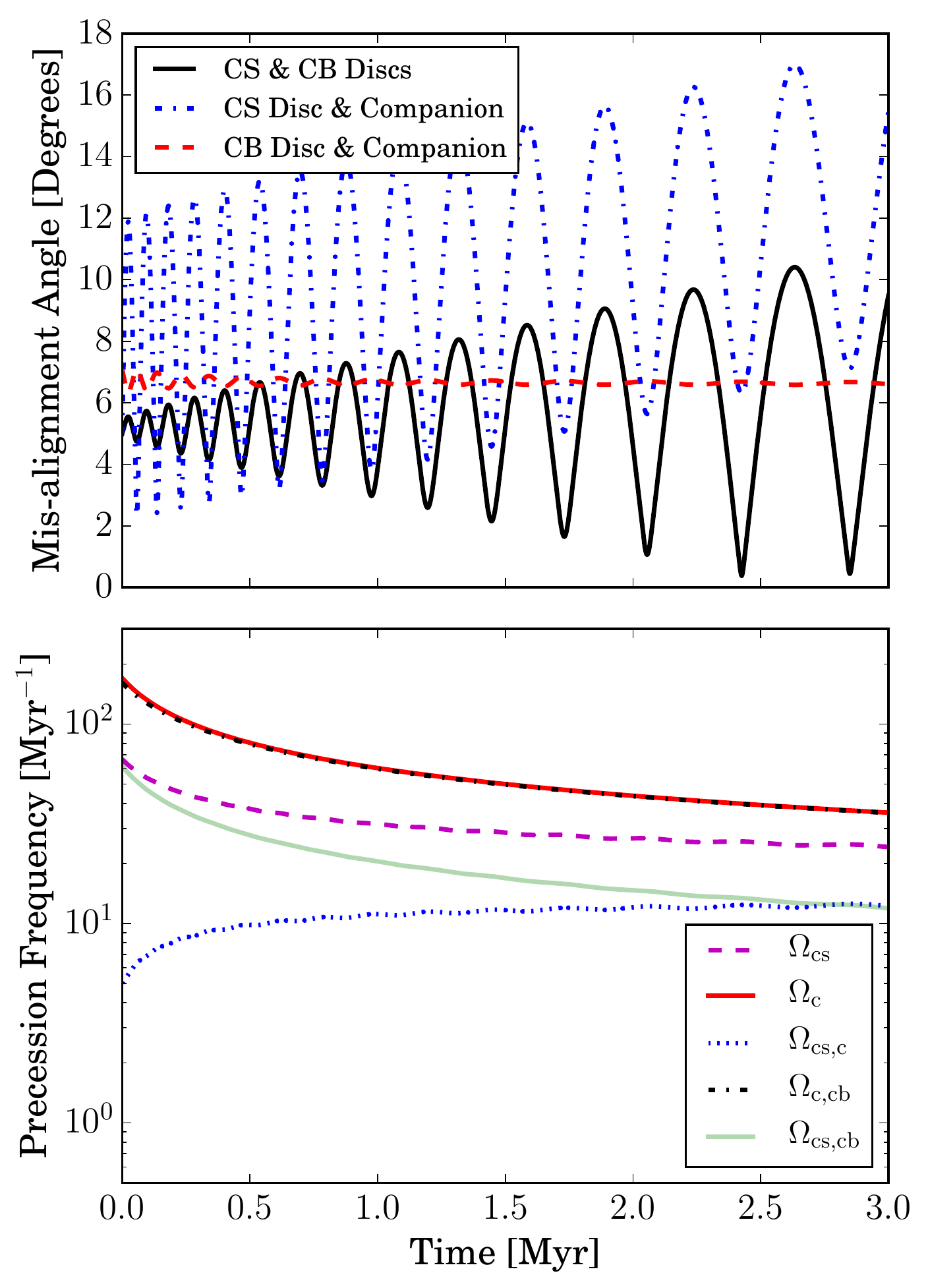}
\caption{The same as Figure~\ref{fig:ac_r1_angle}, expect that $f=0.01$ so that $\Omega_{\rm c}$ and $\Omega_{\rm cs}$ to not become comparable during the evolution of system.}\label{fig:run1_no_res}
\end{figure}

Finally, if the companion does not accrete in this scenario, a resonance between the circumstellar disc and companion's precession never occurs and no large misalignment is generated. This is demonstrated in Figure~\ref{fig:run1_no_res} where we drop the accretion efficiency to $f=0.01$, so that the companion only reaches a mass of $\sim 1.5$~M$_J$ after 3~Myr. A similar outcome is obtained by beginning the evolution with much less mass in the outer disc, such that the companion can not become massive enough to permit resonance crossing. In general, we do not expect misalignments to be generated in planet hosting protoplanetary discs, this conclusion is specifically relevant to whether or not our mechanism operates in ``tranisition'' discs.

\subsection{Migrating Companion}\label{sec:migrate}

In this ``migrating companion'' scenario we consider a companion of fixed mass that has formed in the outer regions of the disc, and is now migrating to smaller separations. We prescribe the migration rate as:
\begin{equation}
\frac{\dot{a}_c}{a_c}=-\frac{1}{t_{\rm mig}}
\end{equation}
where the migration time-scale ($t_{\rm mig}$) is a free parameter. The truncation radii of the discs then evolve according to Equations~\ref{eqn:trun1} \& \ref{eqn:trun2}. For simplicity we do not allow the discs to deplete due to viscous accretion, instead the total mass in the circumstellar and circumbinary discs remains fixed, but the individual mass of each disc evolves in response to the migrating companion. For a disc with a $\Sigma\propto R^{-1}$ surface density profile, the disc mass scales linearly with its outer radius. Thus:
\begin{equation}
\frac{{\rm d}M_{\rm cs}}{{\rm d}t}=M_{\rm cs}\left(\frac{\dot{R}_{\rm cs}}{R_{\rm cs}}\right)=-\frac{M_{\rm cs}}{t_{\rm mig}}
\end{equation}   
and, by mass conservation:
\begin{equation}
\frac{{\rm d}M_{\rm cb}}{{\rm d}t}=-\frac{{\rm d}M_{\rm cs}}{{\rm d}t}=\frac{M_{\rm cs}}{t_{\rm mig}}
\end{equation}
\subsubsection{Results of numerical integrations}

Again we motivate our choice of initial parameters by those similar to the observed systems: $M_*=2\,$M$_\odot$, $a_{\rm c}(t=0)=200$~AU, $M_{\rm c}=0.2\,$M$_\odot$, $M_{\rm cs}(t=0)=0.1\,$M$_\odot$, $M_{\rm cb}(t=0)=0.05\,$M$_{\odot}$ and $R_{\rm out}=1.2R_{\rm in}(t=0)$, with a migration time of $t_\nu=7.5\times10^{5}\,$years. The outer edge of the circumstellar and inner edge of the circumbinary discs are again set using Equations~\ref{eqn:trun1} \& \ref{eqn:trun2}.

\begin{figure}
\centering
\includegraphics[width=\columnwidth]{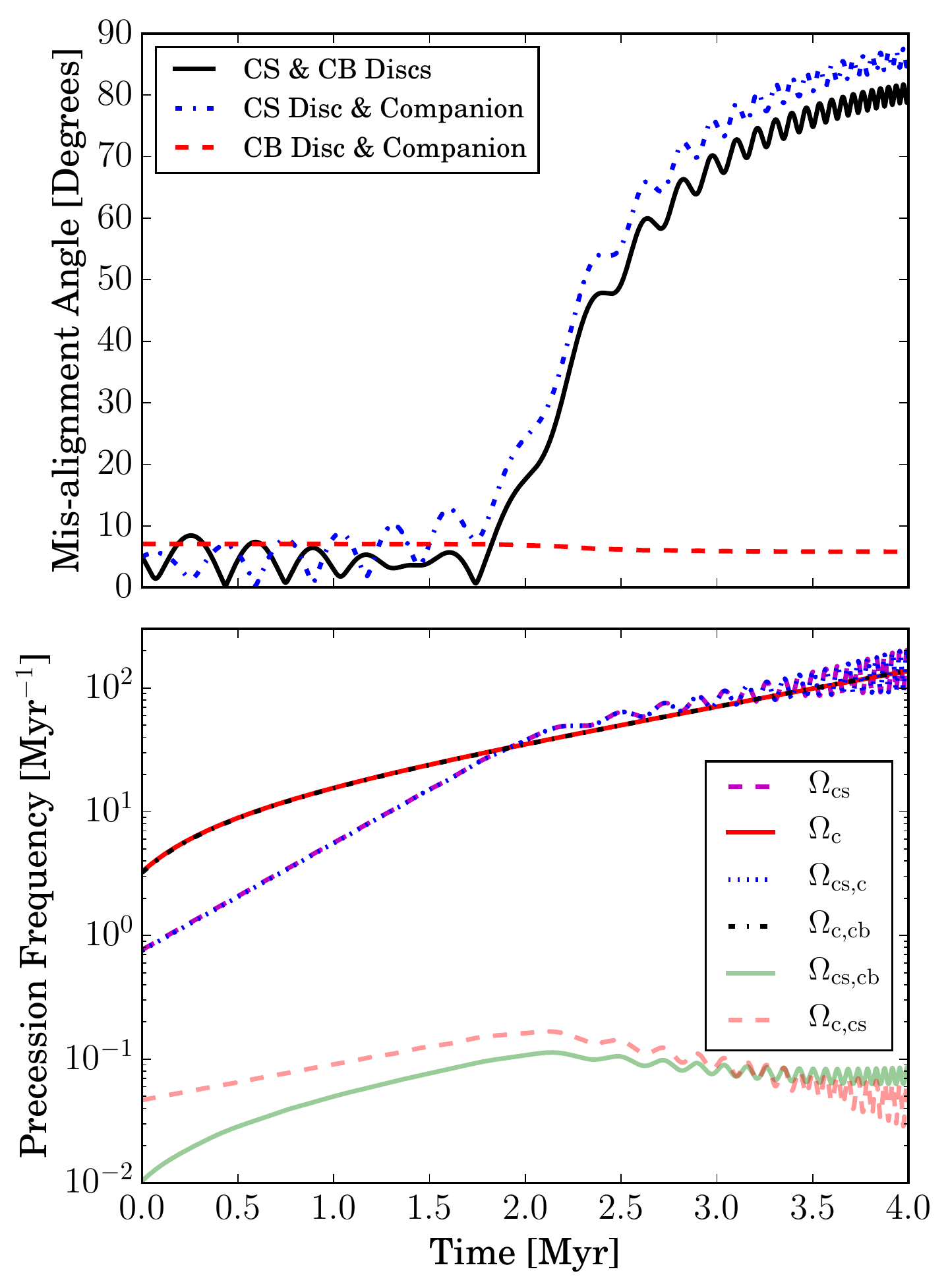}
\caption{The top panel shows the angles between the angular momentum vectors of the circumstellar disc, circumbinary disc and the companion as a function of time. The bottom panel shows the precession frequencies of the circumstellar disc and the companion as well as the indvidual components. These plots are for the case of a migrating companion, starting at 200~AU and migrating inwards with a timescale of $(\dot{a}/a)^{-1}=7.5\times10^{5}$~years.}\label{fig:run2_evol}
\end{figure}

\begin{figure}
\centering
\includegraphics[width=\columnwidth]{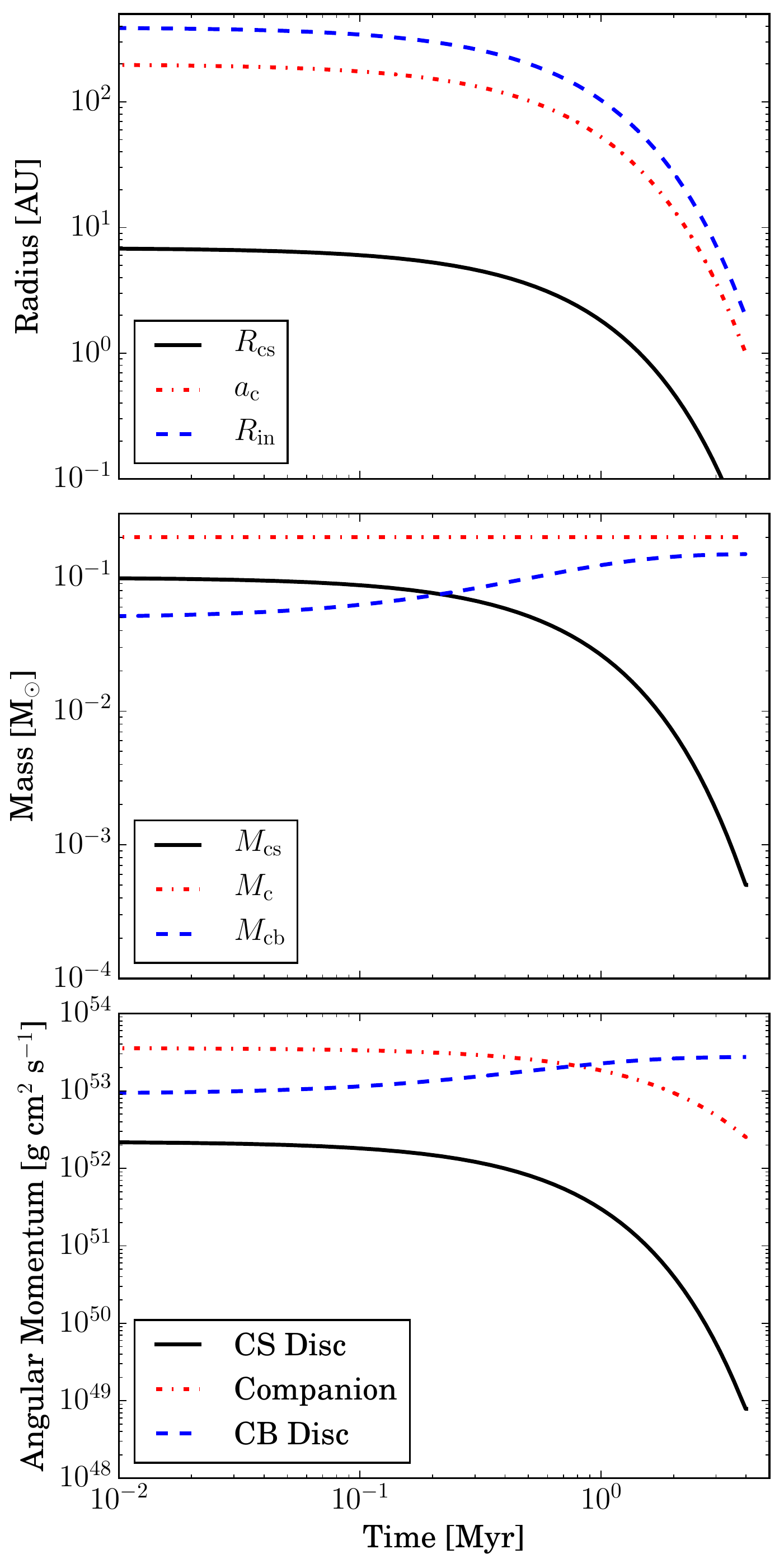}
\caption{Evolution of the disc/companion system as a function of time in the migrating companion scenario. The top panel shows the truncation radii of the circumstellar and circubinary discs. The middle panel shows the evolution of the masses of the circumstellar disc, companion and circumbinary disc and finally the bottom panel shows the evolution of the magnitdue of the  angular momenta of the circumstellar disc, companion and circumbinary disc. The evolving system shown is for the migrating companion scenario depicted in Figure~\ref{fig:run2_evol}.}\label{fig:run2_prop}
\end{figure}

\begin{figure*}
\centering
\includegraphics[width=\textwidth]{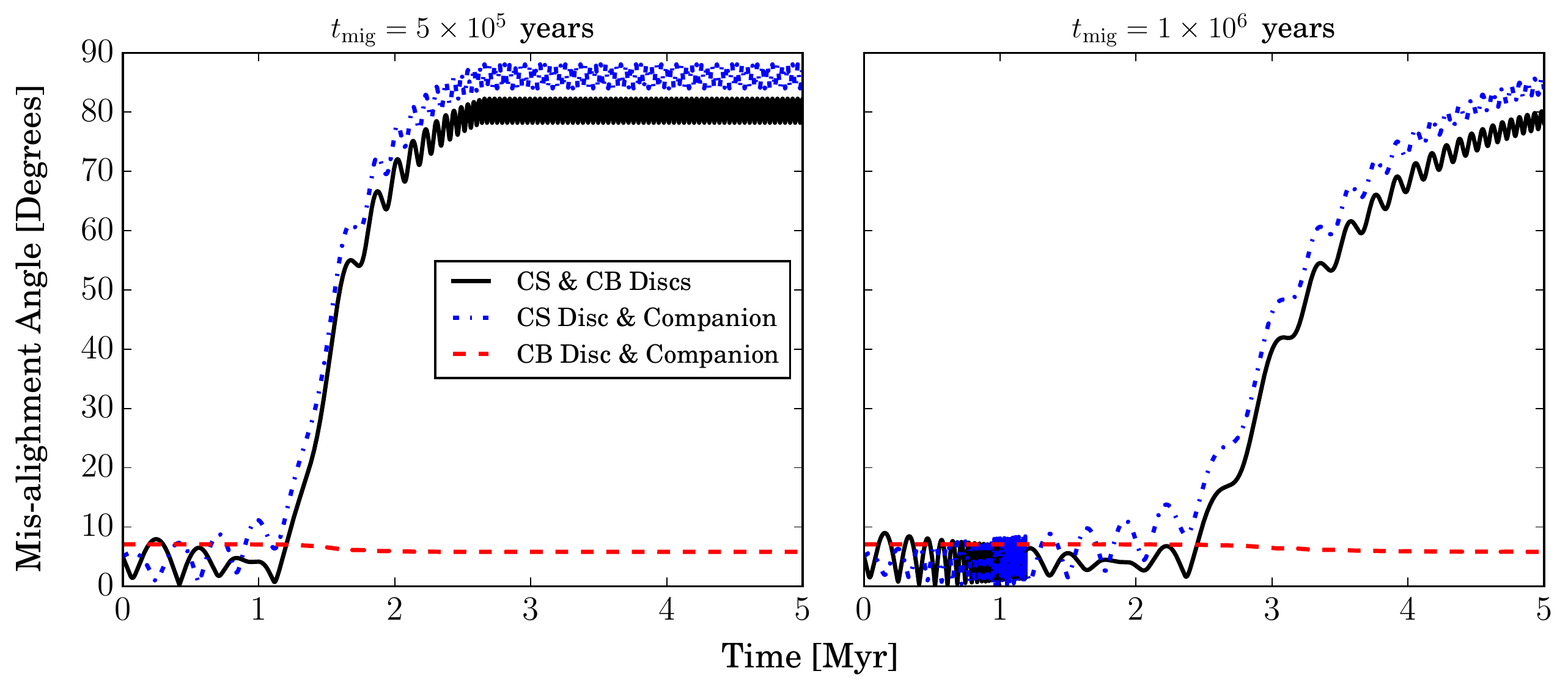}
\caption{The evolution of the misalignment angle for the migrating companion scenario. The system is the same as in Figure~\ref{fig:run2_evol} and Figure~\ref{fig:run2_prop}, except  $t_{\rm mig}=5\times10^{5}$~years in the left panel and $t_{\rm mig}=1\times10^6$~years in the right panel.}\label{fig:run2_vary} 
\end{figure*}

Figure~\ref{fig:run2_evol} shows the evolution of the misalignment angles and precession frequencies for the migrating companion scenario. 
The evolution of the system components are shown in Figure~\ref{fig:run2_prop}.
We see that after about $2\,$Myr the  precession frequencies of the circumstellar disc and companion cross, by this time the companion has migrated to $\sim10\,$AU. Furthermore, the circumbinary disc now contains more mass than the circumstellar disc  and dominates the angular momentum budget of the system. The resonance crossing again causes a large misalignment angle between the two discs ($\sim$70-80 degrees). This evolution is similar to that of the accreting companion scenario discussed in Section~\ref{sec:evolve_discs}, where the secular resonance is dominated by $\Omega_{\rm cs,c}\sim \Omega_{\rm c,cb}$. Again we find that the circumbinary disc and the companion align slightly, meaning the circumstellar disc and companion are also highly misaligned. However, unlike the accreting companion scenario, as the companion continues to migrate to smaller separations, the precession frequency of the circumstellar disc gets shorter  as its outer edge and mass becomes smaller. In Figure~\ref{fig:run2_vary}, we show how varying the migration time between $5\times10^{5}$~years (left panel) and $1\times10^6$ years (right panel) effects the result. The faster migration time leads to a sharper change in the misalignment angles. Similar to the canonical example, after a several Myrs of evolution, the two discs are misaligned by $>70$ degrees.

\subsection{Massive companions and low-mass discs}

Towards the end of the disc's evolution most of the angular momentum of the system will be contained in the binary companion. In this case the companion's orbit will remain unchanged, but could result in nodal precession of the circumstellar and circumbinary discs (Section 3.2). In order to briefly explore this case we consider the migrating companion scenario as described in Section~\ref{sec:migrate}, but reduce the mass of both discs to $5\times10^{-3}\,$M$_\odot$, while keeping the companion's mass to $0.2\,$M$_\odot$. While it is of course questionable whether the companion could still migrate rapidly, our example is illustrative. 

\begin{figure}
\centering
\includegraphics[width=\columnwidth]{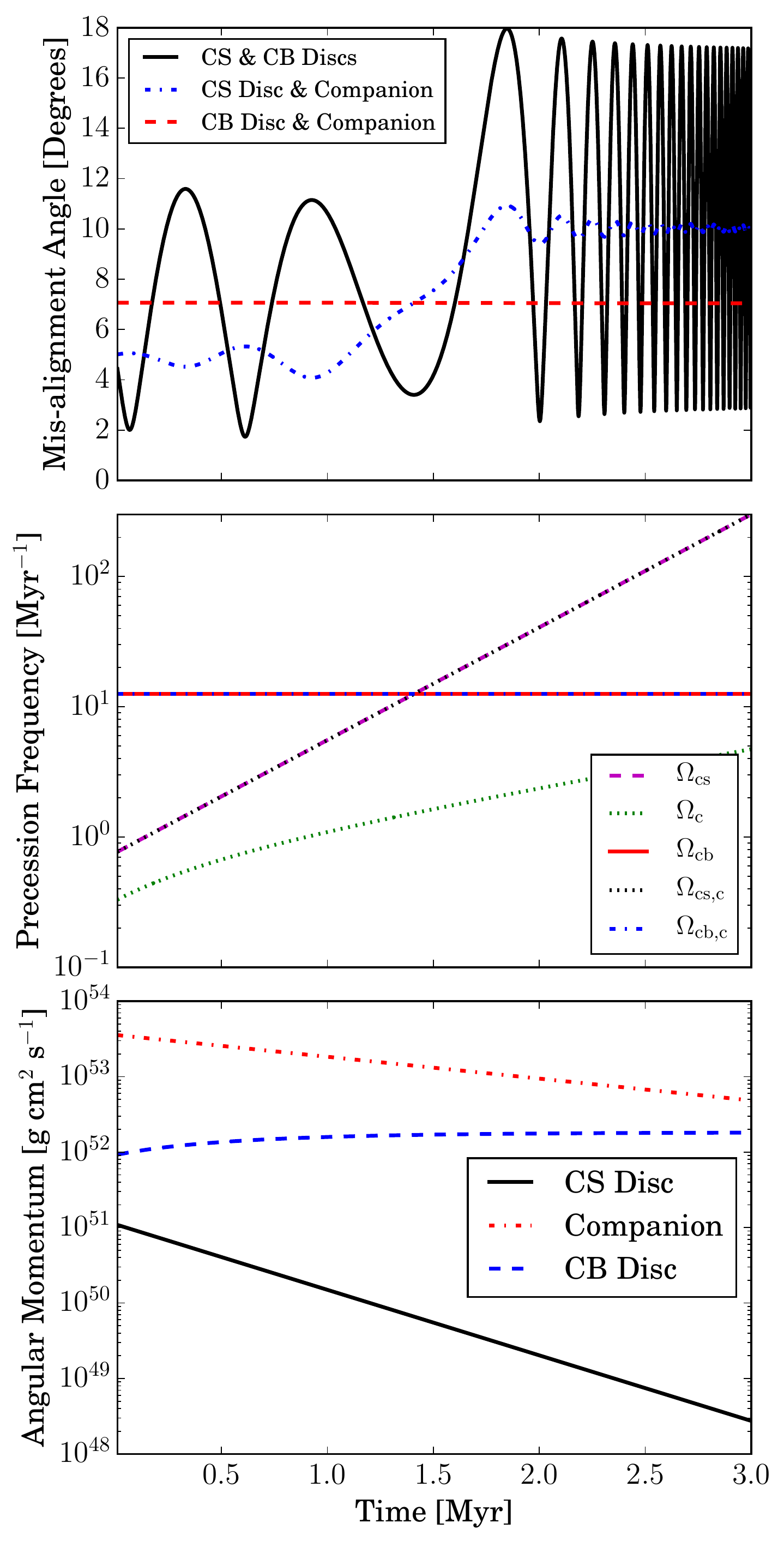}
\caption{The evolution of the misalighnment angles (top), precession frequencies (middle) and angular momenta (bottom) of a system where most of the angular momentum is in the companion.}\label{fig:run3_evol}
\end{figure}
Figure~\ref{fig:run3_evol}, shows the evolution such a system. We see that in this case the companion always dominants the angular momentum budget of the system. The precession frequencies of the circumstellar and circumbinary discs (both of which are completely dominated by driving due to the companion), become equal after about 1.5~Myr of evolution. This causes a modest increase in the misalignment angle (to about 20 degrees) between the circumstellar and circumbinary discs. In general, a ``resonance'' crossing between the circumstellar and circumbinary discs' precession frequencies do not lead to large misalignments from initially aligned configurations.

\section{Discussion}

We have investigated the possibility that secular precession resonances can generate large misalignments between circumstellar and circumbinary discs in young binary systems that have mass ratios of order $0.1$. For realistic systems we find that only resonant interactions between the precession of the circumstellar disc and companion, or between the precession circumstellar and circumbinary disc can occur. 

The resonance between the circumstellar disc and companion is the most robust and the only one that can generate large misalignments ($>60$~degrees) between the two discs, from small (few degree) primordial misalignments.  In a frame precessing with the companion, the circumstellar disc, precesses around an axis almost orthogonal to the disc's angular momentum axis when in resonance, with a precession rate $\Omega_r\approx\sqrt{2}\theta_0\Omega_{\rm cs,c}$, where $\theta_0$ is the small  initial misalignment. In the examples depicted in Figures~1 \& 5,  $\Omega_{\rm cs,c}\sim 50$~Myr$^{-1}$ in resonance, and the rate of change of our system is $\sim 1-5$~Myr$^{-1}$ for both the accreting and migrating companion cases. Thus, the evolution timescale of the system is comparable to the precession time $\Omega_r^{-1}$ (for $\theta_0\sim 0.1$). This means that the circumstellar disc can precess a significant fraction of the way around $\hat{L}_r$ before the system is no longer in resonance. 

Now, we can naturally see why we often get large misalignments if resonance crossing occurs. Young systems will always evolve on $\sim$Myr timescales in the radial range of 10-100~AU. For a companion to star mass ratio ($q$) in the range 0.01-0.1, Equations~\ref{eqn:Ocsc}, \ref{eqn:Or2} \& \ref{eqn:trun1} tell us $\Omega_r\sim 0.1q\theta_0\Omega_{\rm K}(a_{\rm c})$ which for the orbital periods found at 10s of AU of a few hundred years, we get precession timescales about $\hat{L}_r$ of $\sim$Myr. Therefore, it is not surprising in the case of high mass-ratio binaries with separations of order 10s of AU resonance crossing leads to large misalignments. Finally, we note that while it is easy to construct initial and evolutionary parameters that cause our systems to undergo resonance, it is also easy to find examples that do not (see Figure~\ref{fig:run1_no_res}). Therefore, depending on their parameters, real systems may or may not experience resonant misalignment excitations during their lifetimes.



\subsection{Assumptions}\label{sec:assumptions}

In this work, we have made several simplifications in order to make the problem simple and to reveal the basic dynamics. The two most important assumptions are that we ignore explicit angular momentum conservation in our evolving system and we treat both discs as rigid plates.

In the construction of our evolving system we have followed previous work and made no effort to ensure that our disc-companion system explicitly conserves angular momentum. In the case of the accreting disc system, we do not allow the circumbinary disc to expand as it accretes, or lose mass (and angular momentum) in a photoevaporative wind. In the case of viscous accretion with $R_{\rm out}\gg R_{\rm in}$, we would actually expect $R_{\rm out}$ to expand such that the circumbinary disc maintains approximately constant angular momentum in the absence of mass-loss. Since the resonance of interest between the circumstellar disc and companion occurs when most of the angular momentum is in the circumbinary disc, this assumption does not have any important consequences on our calculations. 

In our models, we have implicitly assumed that as  disc material moves from the circumbinary disc to the circumstellar disc (in the case of the accreting system) or from the circumstellar disc to the circumbinary disc (in the case of the migrating companion) it joins the new disc's orbital plane. In reality gas parcels may join the new disc in its original orbital plane. The interaction of material flowing between the two discs is mediated by the companion and it is still uncertain how much angular momentum is absorbed by the companion (and in what orbital plane) during this interaction. However, since we are interested in starting from systems that are initial close to alignment, which then approach a resonance, this assumption should not significantly effect the resonant misalignment excitation. In the case of the accreting system, after the large misalignment has occurred, one would expect accretion from the circumbinary disc onto the circumstellar disc to possibly re-align the two discs on of order Myr timescales (this is similar to the alignment of a binaries orbit which is accreting from a misaligned circumbinary disc discussed by \citealt{Foucart2013,Foucart2014}). Therefore, we imagine large misalignments should readily be observable after the resonance has occurred, even if accretion does slowly realign the discs afterwards. 

In all our models we treat the discs as rigidly precessing plates. In general the discs are warped and twisted \citep{Papaloizou1983}.
 Warp propagation in accretion discs occupies two distinctly different regimes \citep[e.g.][]{Papaloizou1995}, when the viscous $\alpha$ parameter is larger than the aspect ratio ($H/R$) of the disc, the warps propagate diffusively; however, if $\alpha$ is smaller than the aspect ratio the warp is wave-like and propagates with a speed approximately half of the sound speed. 
With aspect ratios of order $0.1$ in protostellar discs at tens of AU \citep[e.g.][]{Chiang1997} and observationally inferred viscous ``alphas'' at least an order of magnitude smaller \citep[e.g.][]{Hartmann1998}, protoplanetary discs are in the wave-like warp regime \citep[e.g.][]{Lubow2000}, where bending waves propagate at approximatively half of the sound speed \citep[e.g.][]{Papaloizou1995}. On timescales longer than the warp propogation time the disc evolves towards and precesses as a rigid body \citep[e.g.][in the linear regime]{Foucart2014}. Since our systems evolve on the long, secular evolutionary timescales,  much longer than the warp propagation time over hundreds of AU, the rigid planet approximation is likely to valid.  

Finally, even though the rigid plate approximation is good, the discs are still likely to maintain a small warp, such a warp can viscously dissipate the mutual inclination between the disc's and companion's orbital planes. This viscous damping rate is highly uncertain (e.g.\citealt{Ogilvie2013} and see discussion in \citealt{Lai2014}); however, \citet{Foucart2014} suggest that large misalignments ($>20\,$degrees) can be maintained for timescales similar to a disc's lifetime. 

\subsection{Observational implications}

One of the motivations for our study was the observed misalignment between the circumstellar and circumbinary discs in HD~142527. HD~142527 is a well known gapped disc with a dust cavity extending from a few tens of AU to $\sim$100~AU \citep[e.g.][]{Fukagawa2006}. Scattered light imaging of the circumbinary disc at NIR wavelengths \citep{Casassus2012,Canovas2013,Avenhaus2014}, 
were interpreted by \citet{Marino2015} as a tilt of $70\pm5$ degrees between the inner (circumstellar) and outer (circumbinary) discs. \citet{Biller2012} discovered a low-mass companion in the cavity of HD~142527 which is now known to be a M dwarf with a mass of order $0.1-0.2\,$M$_\odot$ \citep{Lacour2016} and hence HD~142527 is a high mass-ratio binary system with a mass ratio $q=0.05-0.1$. Therefore, the secular resonance model presented in this work is a likely explanation for the observed misalignment in HD~142527. The precession resonance could have been triggered due to either the companions migration to smaller separations, or accretion of mass onto the companion from the disc, or some combination of both. 


There are several other well known gapped discs which show evidence for large misalignments: HD~100453 \citep{Benisty2016} exhibits a similar misalignment between the inner and outer discs of $\sim 70$ degrees, and HD~135344B \citep{Stolker2016} shows a smaller misalignment of $\sim 22\,$ degrees. No companion has been detected in either system to date. The misalignment in HD~100453 could have been generated by resonance crossing and such a scenario would imply a low-mass $\sim 0.01-0.1\,$M$_\odot$ companion residing in the cavity that is aligned with the outer disc. The misalignment of HD~135433B could have be generated by this mechanism and is in the process of realigning due to accretion, or viscous dissipation. We can probably rule out a resonance between the inner and outer disc in the case of HD~135433B as it requires the companion to dominate the angular momentum budget of the system. HD~135433B has a particularly massive and extended outer disc \citep[e.g.][]{Andrews2011}, which would require the companion to be solar mass or above to induce a large misalignment.   

A precession resonance between the circumstellar disc and companion, starting from a nearly aligned configuration will keep the companion close to alignment with the circumbinary disc. Small arc analysis of approximately 2~years of observations of the companion in HD~142527 has provided weak constraints on the orbital elements of the companion \citep{Lacour2016}. These constraints show the companion is on an eccentric orbit, but the orbital inclination is more uncertain. Lacour et al (2016) suggest (see their Figure~8) that in order for the companion's apocenter to correspond to the inner edge of the circumbinary (outer) disc, it is likely that the companion is aligned with the inner disc. However, we caution that the apocenter distance is not where a companion would truncate the disc \citep[e.g.][]{Artymowicz1994,Miranda2015}. Companions with $q\sim0.1$ are in-between the regime at low-masses where they truncate the disc at several Hill radii \citep[e.g.][]{Crida2006} and comparable mass ratios where they truncated the disc at a distance of a few times the companions separation \citep[e.g.][]{Miranda2015}. Furthermore, the companion will truncate the gas disc, whereas the disc's inner edge is measured from dust emission. It is well known that the gas extends further in than the gas in HD~142527 \citep[e.g.][]{Casassus2015a}. Therefore, it is plausible that the companion is aligned with the outer disc and still play a role in carving the circumbinary disc. For example, applying the results of \citet{Artymowicz1994} indicates an aligned binary with mass ratio $q\sim0.1$, an eccentricity of $\sim 0.5$ will truncate the circumbinary disc between $3-4$ times the binary separation. HD~142527's binary separation is $20^{+17}_{-10}$~AU and eccentricity is $0.5\pm0.2$; thus, the companion can orbit in the same plane as the circumbinary disc and still truncate the disc's dust component at $\sim 100$~AU. Further monitoring of the companion should be able to determine its orbital plane, and test the resonant scenario presented here.

\section{Summary}
We have shown that secular precession resonances can operate in protostellar gapped/binary discs, which can result in a misalignment between the circumstellar (inner) and circumbinary (outer) discs. For some cases, the generated misalignments can be large ($\gtrsim 60$~degrees) resulting from crossing a precession resonance between the circumstellar disc and a low-mass companion (or massive planet) as the system evolves. Our main results are summarised below:
\begin{enumerate}
\item We identify two secular precession resonances in gapped/binary disc systems. First, a resonance between the precession of the circumstellar and circumbinary discs. Second, a resonance between the precession of the circumstellar disc and a low-mass companion residing in a gap between the circumstellar and circumbinary disc.\\

\item The resonance between the circumstellar and circumbinary disc cannot lead to large misalignments as both discs precess independently, even in ``resonance''. \\

\item The resonance between the circumstellar disc and companion can lead to significant misalignments, even from systems that are close to being co-planar initially, provided that the resonance crossing occurs on the right timescale. This typically requires that the companion has a mass $\sim 0.01-0.1\,$M$_\odot$, separation $\sim 10-100\,$AU for an approximately solar mass primary and that the circumbinary disc dominates the angular momentum budget of the system. Such requirements are {\it not} satisfied by giant -- of order Jupiter mass -- planets in ``transition'' discs.  \\

\item In numerical calculations of realistic systems we find misalignments of $\sim 70$~degrees, consistent with those observed in HD~142527 and HD~100453. Our results indicate a secular resonance between a companion and the circumstellar (inner) disc in HD~142527 and HD~100453  occurred previously in their histories and misaligned their two discs. Like the companion present in HD~142527, we would suggest a companion exists in HD~100453 which is either a very massive planet, brown dwarf, or low-mass star.    \\

\item For gapped disc systems that are initially close to being co-planar, the secular resonance mechanism would predict the companion to remain close to co-planar with the circumbinary (outer) disc, but misaligned to the circumstellar (inner) disc. This prediction is testable in HD~142527 with a longer baseline of observations of the companion's orbit. 
\end{enumerate}

\section*{Acknowledgements}

JEO acknowledges support by NASA through Hubble Fellowship grant HST-HF2-51346.001-A awarded by the Space Telescope Science Institute, which is operated by the Association of Universities for Research in Astronomy, Inc, for NASA, under contract NAS 5-26555. DL has been supported in part by NASA grants NNX14AG94G and NNX14AP31G, and a Simons Fellowship from the Simons Foundation.







\bsp	
\label{lastpage}
\end{document}